# CHAPTER 7:

# JOINT INVERSION IN HYDROGEOPHYSICS AND NEAR-SURFACE GEOPHYSICS

Niklas Linde[1] and Joseph Doetsch[2]

[1]Applied and Environmental Geophysics Group, Institute of Earth Sciences, University of Lausanne, 1015 Lausanne, Switzerland;

[2]Swiss Competence Center for Energy Research – Supply of Electricity (SCCER-SoE), ETH Zurich, 8092 Zurich, Switzerland.

**Key points:**

- Structural joint inversions of near-surface geophysical data are often favored

- Time-lapse geophysical data are sensitive to hydrological state variables

- Joint inversion in hydrogeophysics should include flow- and transport modeling

**Abstract**

The near-surface environment is often too complex to enable inference of hydrological and environmental variables using one geophysical data type alone. Joint inversion and coupled inverse modeling involving numerical flow- and transport simulators have, in the last decade, played important roles in pushing applications towards increasingly challenging targets. Joint inversion of geophysical data that is based on structural constraints is often favored over model coupling based on explicit petrophysical relationships. More specifically, cross-gradient joint inversion has been applied to a wide range of near-surface applications and geophysical data types. To infer hydrological subsurface properties, the most appropriate approach is often to use temporal changes in geophysical data that can be related to hydrological state variables. This allows using geophysical data as indirect hydrological observables, while the coupling with a flow- and transport simulator ensures physical consistency. Future research avenues include investigating the validity of different coupling strategies at various scales, the spatial statistics of near-surface petrophysical relationships, the influence of the model conceptualization, fully probabilistic joint inversions, and how to include complex prior information in the joint inversion.

**7.1 Introduction**

The increasing pace of unsustainable man-induced activities and associated threats [*Rockström et al.*, 2009] will, for the foreseeable future, continue to push researchers and practitioners towards environmental problems of growing complexity. Remote sensing and geophysics are playing ever-increasing roles in describing near-surface environments at scales spanning the root zone to major aquifers [*NRC*, 2012]. For instance, critical zone research focusing on "*the heterogeneous, near-surface* *environment in which complex interactions involving rock, soil, water, and living organisms regulate the natural habitat and determine the availability of life-sustaining resources"* [*NRC*, 2001] requires spatially-distributed data. Especially, process understanding in the deep critical zone can be enhanced by geophysical data, as it is largely out of reach for classical methods based on coring and trial pits [*Parsekian et al.*, 2015]. To illustrate the need for multiple data and joint inversion, let us consider the discipline of hydrogeophysics that relies on geophysical data to gain information about hydrological processes or controlling subsurface structures [e.g., *Rubin and Hubbard*, 2005; *Hubbard and Linde*, 2011]. Early hydrogeophysical research often relied on the assumption that geophysical tomograms could be treated as spatially distributed and exhaustive "data". Petrophysical relationships (sometimes estimated through inversion; e.g., *Hyndman et al.*, [2000]; *Linde et al.* [2006b]) were then used to convert these "data" into hydrological properties [e.g., *Rubin et al.*, 1992; *Copty et al.*, 1993; *Cassiani and Medina*, 1997; *Cassiani et al.*, 1998]. Such approaches have been criticized as they often neglect the resolution limitations of geophysical tomograms [*Day-Lewis and Lane*, 2004; *Day-Lewis et al.*, 2005] and for the strong assumptions made about the relationship between the geophysical tomogram and the hydrogeological system of interest [*Linde et al.*, 2006c]. Indeed, using one data type alone is often insufficient to adequately constrain key target variables [e.g., *Linde et al.*, 2006a]. Joint inversion provides formal approaches to integrate multiple data such that the resulting subsurface models and interpretations are more consistent and reliable than those obtained by comparing the results obtained by individual inversion of different data types [e.g., *Doetsch et al.*, 2010b]. *Linde et al.* [2006c] conclude their review on hydrogeophysical parameter estimation approaches by stating that joint inversion is the way forward, but that is still in its infancy. The situation has changed radically in the last 10 years and joint inversion is nowadays widely used.

Most of the joint inversion methodologies have been developed and applied for hydrogeophysical applications and it is hoped that this review can stimulate applications in other related fields such as archeological geophysics and civil engineering geophysics. Archeological prospecting would be an ideal target for joint inversion, because archeological remains cause signatures in multiple geophysical methods (e.g., magnetometry, ground penetrating radar (GPR), electrical resistance tomography (ERT)). Many archeological sites have been surveyed with complementary geophysical methods, which offer the possibility of joint interpretation and data integration [e.g., *Böniger and Tronicke*, 2010; *Keay et al.*, 2009]. To date, integrated analysis has been performed on the 2D and 3D models that result from the individual geophysical surveys using visualization tools, advanced image processing and statistical analysis [*Kvamme*, 2006; *Watters*, 2006]. Joint inversion could here help to shift the integration to the data level and ensure that resultant models honor all available data. In engineering and public safety applications, joint inversion can improve the reliability of feature detection (e.g., unexploded ordnance (UXO) characterization: *Pasion et al.*, [2003]), which could also be used for safety assessment of critical infrastructure such as roads and bridges.

This review focuses on joint inversion of geophysical data, as well as joint inversion of geophysical and hydrological data, in near-surface environments. To enable a comprehensive review, we primarily consider contributions that fulfill the following conditions:

1. At least one type of near-surface geophysical data is considered. We will not consider the vast literature on joint inversion of different types of hydrological data, such as, hydraulic head and tracer test data [*Nowak and Cirpka*, 2006]);

2. At least two different physical or geological subsurface properties are considered. For example, this implies that we will not discuss joint inversion of electrical and electromagnetic (EM) induction methods, as they are both sensitive to electrical resistivity [e.g., *Kalscheuer et al.*, 2010; *Rosas-Carbajal et al.*, 2014]). These type of applications can be handled within the same framework as individual inversions;

3. There is one common objective function that considers the data misfit of all data types simultaneously. We will not discuss inversion of one geophysical data type that uses prior information about property variations [e.g., *Saunders et al.*, 2005] or lithological interfaces [*Doetsch et al.*, 2012] inferred from previously processed or inverted geophysical data. These are very useful approaches, but they do not qualify as joint inversion;

4. We consider joint inversion of dependent data only (e.g., electrical resistances, seismic traveltimes, tracer breakthrough curves). That is, we do not consider geophysical inversion of one data type conditional to interpreted borehole logging data (e.g., fracture zonation by combining permeability estimates based on flowmeter data with crosshole seismic traveltimes [*Chen et al.*, 2006]). Again, such approaches can be very useful, but they do not qualify as joint inversion;

5. The primary depth range is from the centimeter to the kilometer scale and we exclude application areas treated elsewhere in this book volume. Hence, we will not consider mineral exploration [see chapter 8], hydrocarbon exploration [see chapter 9] or geothermal applications [see chapter 10]).

This review is organized as follows. Section 7.2 describes petrophysical approaches and section 7.3 describes structural approaches to joint inversion in near-surface environments. Section 7.4 focuses on hydrogeophysical applications in which hydrological flow- and transport simulations are combined with petrophysical relationships that link hydrological state variables and geophysical properties. Section 7.5 discusses outstanding challenges and section 7.6 provides concluding remarks.

## 7.2. Petrophysical joint inversion in near-surface environments

Petrophysical joint inversion relies on a statistical or theoretically based relationship between different model parameter types (e.g., P-wave velocity and density; electrical conductivity and permeability). One attractive feature of joint inversion by petrophysical coupling is that it allows formulating the inverse problem in terms of the target properties of primary interest (e.g., porosity, permeability, lithology). In this approach that is referred to as lithological tomography, the petrophysical relationships are used to transform the primary property fields into geophysical property fields that the observed geophysical data are sensitive to (see the upper arrows in Figure 7.1). Comparisons of the simulated forward responses and the data are then used to guide further model updates. For more details, we refer to the work by *Bosch* [1999; 2015]. Another option is to impose a direct petrophysical relationship between different geophysical properties (e.g., electrical conductivity, dielectric permittivity; see lower arrow in Figure 7.1a). If the petrophysical relationships include primary hydrological properties (e.g., porosity or permeability) and not hydrological state variables (e.g., pressure, water content, salinity), then there is no methodological difference in including hydrological data in the joint inversion (see the upper three rectangles in Figure 7.1b). In the following, hydrological data should be understood as measurements of hydrological state variables (e.g., hydraulic pressure or salinity).

There are until recently only few examples of petrophysics-based joint inversion in near-surface environments. The likely reason for this is that the approach is mainly feasible when working with global sampling or optimization methods [*Sen and Stoffa*, 2013] that can easily consider non-linear, complex, and uncertain petrophysical relationships. Local inverse formulations (e.g., Gauss-Newton) are very sensitive to errors in the petrophysical model and it is likely that model artifacts will be introduced to compensate for these errors. It is only recently that the near-surface community has adapted global methods and computational constraints will continue to limit their applicability for many applications. For example, *Hertrich and Yaramanci* [2002] used simulated annealing to jointly invert both synthetic and field-based surface nuclear magnetic resonance (SNMR) and vertical electrical sounding (VES) data under the assumption of a common layered 1-D earth structure. Each layer was parameterized in terms of mobile (sensed by SNMR and VES data) and immobile (sensed by the VES data only) water content, fluid resistivity and the cementation exponent in Archie's law. *Jardani et al.* [2010] used an adaptive Metropolis algorithm to jointly invert synthetic seismic and seismoelectric data to infer, in a 2-D setting, permeability, porosity, electrical conductivity, and four different moduli for three zones of known geometry.

We are not aware of work on joint inversions of geophysical and hydrological data that use a petrophysical relationship to relate primary hydrological properties with geophysical properties. Most joint inversions of geophysical and hydrological data rely on petrophysical relationships between geophysical properties and hydrological state variables. Such approaches are discussed in section 7.4.

**7.3. Structurally coupled joint inversion in near-surface environments**

Structurally coupled joint inversions seek multiple distributed models that share common interfaces or have similar model gradients (see Figure 7.1a). They can be grouped into three sub-categories: common layers, common lithological units and gradient-based joint inversion.

*7.3.1. Common layer structure*

It is possible to conceptualize the subsurface by layers of constant (1-D) or variable (2-D and 3-D) thickness and to ignore vertical property variations within layers. The model parameters to infer are thus related to the interface depths between layers and lateral property variations with each layer. This is often a suitable model parameterization for large-scale aquifer characterization in sedimentary settings and it facilitates the integration of unit boundaries identified in borehole logs. The coupling for the joint inversion is here achieved by imposing common layer boundaries [*Auken and Christiansen*, 2004].

*Hering et al.* [1995] jointly inverted VES data together with Rayleigh and Love wave dispersion curves for a layered 1-D model. In a synthetic example it was found that the resulting model parameters were better defined than for individual inversions. *Misiek et al.* [1997] applied this joint inversion approach to two field data sets and reported improved results compared with individual inversions. *Kis* [2002] proposed a joint inversion of VES and seismic refraction traveltimes that she applied to both synthetic and field-based data. The seismic and electrical properties were assumed constant within each layer and they shared common interfaces in 2-D by using a common parameterization in terms of Chebyshev polynomials. *Wisén and Christiansen* [2005] derived 2-D layered models based on 1-D modeling for both synthetic and field-based VES and surface wave seismic data. The models were made laterally continuous by penalizing lateral gradients of layer properties and interface locations. The electrical resistivity and shear wave velocity models were coupled by assuming similar layer interfaces. This coupling led to a better-constrained shear wave velocity model and a good agreement of interfaces identified with lithological drill logs. *Juhojuntti and Kamm* [2010] further extended this approach to 2-D joint inversion of both synthetic and field-based seismic refraction and ERT data. The resulting interfaces agreed well with independent information (cone-penetration tests, drillings, a reflection seismic section). *Santos et al.* [2006] jointly inverted VES and gravity data for a 2-D layered model (the VES data were simulated using a local 1-D model). The density of each layer was assumed known and simulated annealing was used to determine the layer resistivities and thicknesses. Synthetic test models indicated that the joint inversion provided a more stable and reliable parameter estimation than individual inversions. A field example related to water prospection provided results in agreement with local geology. *Jardani et al.* [2007] used simulated annealing to jointly invert field-based self-potential and apparent electrical conductivity data (EM-34). Their model consisted of two layers with an irregular interface. Each layer was characterized by an electrical resistivity and an apparent voltage-coupling coefficient. The modeling was 1-D, but lateral constraints were imposed to obtain a map view of the depth to the interface, which was used to identify sinkholes. *Moghadas et al.* [2010] jointly inverted full-waveform data from synthetic off-ground zero-offset GPR and EM data to determine the electrical conductivity and permittivity of two soil layers, as well as the thickness of the upper layer. A Bayesian formulation to the joint inversion provided the best results and they found that sequential inversion alternatives could lead to biased estimates. *Günther and Müller-Petke* [2012] jointly inverted SNMR and VES data using a Marquardt-type Gauss-Newton inversion. In addition to the thickness of each layer, the algorithm provided the porosity, decay time, and electrical resistivity of each layer. In a field application, they found that the inclusion of SNMR data resulted in an improved lithological description of the subsurface compared with inversion of VES data alone.

### 7.3.2. Common lithological units

The layered model parameterization allows for variable layer thicknesses, but the number of layers is the same throughout the model domain and each layer has an infinite lateral extent. This is a questionable assumption in geological settings characterized by finite-size bodies, for example, channel forms or clay lenses. In such cases, it is better to parameterize the subsurface in terms of lithological units and to assume that the internal heterogeneity within each unit is negligible compared to the contrast with neighboring units. The main challenge with this formulation is the need for an adaptive model parameterization.

*Musil et al.* [2003] used discrete tomography based on mixed-integer linear programming to jointly invert synthetic crosshole seismic and radar data to locate air and water filled cavities. Inequality constraints forced model properties to the expected values (rock, air- and water-filled cavities) with discrete jumps at the interfaces. *Paasche and Tronicke* [2007] proposed a cooperative inversion, in which the results of individual inversions (synthetic crosshole GPR and P-wave traveltimes in their examples) underwent a fuzzy c-means clustering step after each iteration step. The resulting zonal model was subsequently used as the starting model for a conventional individual inversion, and so on. This allows information sharing from different data types, even if it is strictly speaking not a joint inversion. *Linder at al.* [2010] extended this approach to three data types (field-based crosshole GPR, P-wave and S-wave traveltimes).

Considering a synthetic test case, *Hyndman et al.* [1994] inverted for lithological models by simultaneously minimizing the misfit between simulated and observed seismic traveltimes and tracer test data. A standard seismic traveltime inversion was first carried out to obtain an initial seismic velocity model. A separation velocity between homogeneous high and low velocity zones was then sought that minimized the traveltime residuals. Under the assumption that this zonal model was also representative of permeability, the tracer test data were inverted for a constant permeability in each zone. In the next step, the seismic velocity model was updated by performing a new seismic inversion with the zonal model as starting model. The separation velocity was this time determined by simultaneously minimizing the misfit in seismic traveltimes and the tracer test data. These steps were repeated multiple times. In a field application, *Hyndman and Gorelick* [1996] extended this approach from 2-D to 3-D, into three different lithology types, and including hydraulic data as well. The resulting 3-D lithological model provides one of the most convincing hydrogeophysical case studies to date. *Cardiff and Kitanidis* [2009] developed an extended level set method to obtain zonal models with geometries (shape and location) being dependent on all data types. A synthetic 2-D example was used to demonstrate joint inversion of seismic traveltimes and drawdown data from pumping tests at steady state. The final zonal model was vastly improved compared with individual inversions (see Figure 7.2), both in terms of shape and parameter values of the sand and clay inclusions in the gravel background. For a synthetic text case, *Aghasi et al.* [2013] presented a level set method to infer a contaminant's source zone architecture in 3-D by joint inversion of ERT and contaminant concentration. Level sets have the distinct advantage that the interfaces between different geological units are discontinuous. In many settings, this is necessary to reproduce hydrogeological properties and responses.

Conceptually similar to the case of common lithological units is the detection of man-made structures and objects in the subsurface. Examples could be archeological remains hidden in sediments or buried metallic objects, such as UXO. *Pasion et al.* [2003] developed a methodology for jointly inverting magnetic and time-domain electromagnetic (TDEM) data for characterization of UXO. Due to the known size and property range for UXO and the strong magnetic and electrical contrast to the background, the inversion could be formulated to invert directly for UXO characteristics such as position, orientation and magnetic and electrical properties. Magnetic and TDEM data jointly contributed to resolving position and orientation, while the other parameters were not shared between the methods. In the synthetic example of *Pasion et al.,* [2003], joint inversion were found to improve size and shape estimates of the buried target.

### 7.3.3. Gradient constraints

The most popular model parameterization strategy in geophysical inversion is to use a very fine model discretization that remains fixed during the inversion process. The inverse problem is formulated to maximize the weight assigned to model regularization constraints that quantify model structure provided that the data are fitted to the expected noise level. This Occam-style inversion can easily be solved using a least-squares formulation and it leads to a minimum-structure model, in which all resolved features are necessary to explain the observed data [*Constable et al.*, 1987]. The resulting images are smoothly varying fields that are visually very different from the discontinuous property fields that are discussed in sections 7.2.1 and 7.2.2. Gradient-based joint inversion has played an important role in popularizing joint inversion as they are easily implemented in existing Occam-style inversion algorithms. Most published papers rely on the cross-gradients function introduced by *Gallardo and Meju* [2003]. This approach is valid when changes in the physical properties of interest are aligned (i.e., gradients are parallel, anti-parallel or the gradient is zero for one property). This is a reasonable assumption when a single lithological property dominates the subsurface response (e.g., porosity) or when changes in state variables are also large across lithological units. The cross-gradients constraints are invalid (in its standard formulation) when there are important uncorrelated changes in lithological and state variables [e.g., *Linde et al.*, 2006a].

The original cross-gradients joint inversion formulation [*Gallardo and Meju*, 2003] was first applied to 2-D surface-based near-surface geophysical field data [*Gallardo and Meju*, 2003, 2004, 2007]. A slightly modified formulation was introduced by *Linde et al.* [2006a] and applied to joint 3-D inversion of field-based crosshole GPR and ERT data. This approach was later adapted to joint inversion of synthetic and field-based crosshole seismic and GPR data [*Linde et al.*, 2008], to synthetic and field-based time-lapse crosshole ERT and GPR data [*Doetsch et al.*, 2010a] and to synthetic and field-based three-method joint inversion in combination with classification, clustering, and zonal joint inversion [*Doetsch et al.*, 2010b]. *Bouchedda et al.* [2012], *Karaoulis et al.* [2012], *Hamdan and Vafidis* [2013] and

*Revil at al.* [2013] presented related applications. This suite of real-world case-studies based on different data types and geological settings suggest that cross-gradients joint inversion is presently one of the most robust approaches to joint inversion of near-surface geophysical field data. The resulting models have higher resolution and cross-property plots are more focused than for individual inversions (see Figure 7.3). The focused scatter plots enable visual and automatic clustering that facilitate geological interpretations and petrophysical inference [e.g., *Gallardo and Meju*, 2004; *Doetsch et al.*, 2010b].

*Günther and Rücker* [2006] introduced an alternative gradient-based joint inversion approach that they applied to synthetic 2-D seismic refraction and ERT data. They used the model gradients in one model to locally scale the regularization constraints in the other model, and vice versa. If one property displays large spatial changes at one location, this approach helps to introduce larger changes in the other property field by locally decreasing the regularization weights. This approach has not been used extensively and any added value with respect to cross-gradients constraints remains to be demonstrated.

*Lochbühler et al.* [2013] presented the first joint inversion of synthetic and field-based geophysical and hydrological data based on gradient-constraints. The cross-gradients function was used to impose structural similarities between radar slowness and the logarithm of the hydraulic diffusivity (or the permeability field) when inverting crosshole GPR and hydraulic tomography (or tracer) data. Similar to previous applications, the joint inversion provided models with higher resolution and cross-property plots were less scattered than for individual inversions.

**7.4. Coupled hydrogeophysical inversion**

One popular approach to jointly invert geophysical and hydrological data is to link hydrological state variables to geophysical properties. This approach has primarily been applied to transient hydrological phenomena using time-lapse geophysical data. Typical examples include water flow in the unsaturated (vadose) zone and salt tracer movement in saturated aquifers. Classical time-lapse geophysical inversion suffers from resolution limitations. In tracer experiments, this might lead to the inferred plumes being unphysical [e.g., loss of mass; *Day-Lewis et al.*, 2007]. When considering geophysical data within a hydrological inversion context, all results are, by construction, in agreement with mass conservation laws and other constraints imposed by the hydrological model. Another advantage is that petrophysical relationships related to a perturbation in a hydrological state variable (e.g., salinity, water content) are much better constrained than for primary properties. In time-lapse inversions it is expected that only state variables change with time, whereas primary properties and petrophysical parameter values remain unchanged. This reduces the number of unknown petrophysical parameters that need to be assigned or inverted for.

For coupled hydrogeophysical inversions it is challenging to provide a general implementation and parameterization for a wide range of applications, hydrological settings and geophysical data, and most published works are specific to certain settings (e.g., water flow in unsaturated soil) and geophysical data types. It is also often necessary to assume that certain properties are constant (e.g., porosity) when inverting for the spatial variations of others (e.g., permeability). Furthermore, petrophysical parameters (e.g., the cementation exponent in Archie's law) are often assumed constant throughout the study area and the consequences of such assumptions are seldom addressed in detail.

*7.4.1. Petrophysical coupling applied to vadose zone hydrology*

Variations in water content have a clear geophysical signature, for example, in terms of electrical permittivity and resistivity. This implies that time-lapse geophysical data can be used to monitor vadose zone processes, but also to constrain subsurface architecture and hydrological properties within a coupled hydrogeophysical inversion. The most commonly used geophysical data have been GPR and ERT data acquired with crosshole or surface acquisition geometries.

*Kowalsky et al.* [2004] pioneered coupled hydrogeophysical inversion by linking an unsaturated flow simulator and a GPR forward solver. In their approach, the soil hydraulic properties were described by the van Genuchten parameterization [*van Genuchten*, 1980] that relates permeability to water content in partially saturated media. Water content was related to permittivity and GPR velocities using the complex refractive index model [CRIM, *Roth et al.*, 1990]. *Kowalsky et al.* [2004] used a pilot point parameterization to generate multi-Gaussian fields with a reduced set of model parameters. In a synthetic experiment, the crosshole GPR measurements were jointly inverted with water content values that were available along the boreholes. The coupled inversion improved the estimates of permeability and its spatial variation compared with considering the water content data only. It was also found that permeability could be very well resolved, when all other soil hydraulic properties and the petrophysical parameters were perfectly known. In addition, they retrieved soil hydraulic and petrophysical properties under the assumption of homogeneous distributions. *Kowalsky et al.* [2005] inverted multi-offset crosshole GPR field data with conditioning to point information of water content (estimated from Neutron probe measurements). They simultaneously estimated homogeneous petrophysical parameters and the heterogeneous distribution of permeability.

Including GPR data improved permeability estimates in a synthetic example (see Fig. 7.4), which led to a better prediction of water content. *Kowalsky et al.* [2005] concluded that it was crucial to infer petrophysical parameter values within the inversion procedure, as incorrect assumptions might otherwise compromise the permeability estimates. Nevertheless, it is important to acknowledge that assuming homogeneous petrophysical parameters (as done herein) could also bias the permeability estimates. *Finsterle and Kowalsky* [2008] inverted for geostatistical parameters of unsaturated soil. Using filtration rates, water content and GPR traveltimes, they were able to simultaneously invert synthetic data for three homogeneous hydraulic parameters, two petrophysical parameters and three geostatistical parameters. The permeability distribution parameterized by pilot points was also part of the inversion.

*Cassiani and Binley* [2005] inverted for soil hydraulic properties (Mualem - van Genuchten parameters) of different layers by coupling a Richards' equation solver to a Monte Carlo sampler and linked water content to zero-offset GPR data through a known petrophysical relationship. In a field example, *Cassiani and Binley* [2005] found that unsaturated flow parameters were not individually constrained by their data and that measurements under dynamic conditions would have been needed. They also stressed the importance of acquiring independent geological information. *Scholer et al.* [2011, 2012] used a Markov chain Monte Carlo (MCMC) inversion approach to study the influence of prior information on estimated soil hydraulic properties. They found when using both synthetic and field-based test cases that the geophysical data alone contained valuable information, but that significantly better results were obtained when using informative prior distributions that include parameter correlations.

All of the approaches presented above relied on coupled hydrogeophysical inversion based on crosshole GPR traveltime data to provide information in-between the boreholes. A number of papers have addressed the coupling of unsaturated flow modeling (e.g., based on Richards' equation) with surface GPR data. *Lambot et al.* [2006] and *Jadoon et al.* [2008] presented numerical experiments with off-ground GPR monitoring data to constrain the hydraulic properties of the topsoil and *Busch et al.* [2013] inferred such soil hydraulic properties from both synthetic and field-based surface GPR measurements. These methods could potentially cover larger areas than crosshole applications, although frequent repeat measurements would be needed to retrieve soil hydraulic properties.

The above-mentioned examples highlight the utility of GPR for retrieving soil hydraulic properties. *Looms et al.* [2008] combined 1-D flow simulations with ERT and GPR data to invert for permeability. The ERT and GPR forward simulators were both linked to the hydraulic simulator through a known and homogeneous petrophysical relationships. *Looms et al.* [2008] showed that five layers were needed to explain their field data, but that only permeability and one additional fitting parameter could be retrieved for each layer.

*Hinnell et al.* [2010] and *Mboh et al.* [2012a] investigated how surface-based ERT monitoring of infiltration tests could improve soil hydraulic property estimation. *Hinnell et al.* [2010] showed in a synthetic example that petrophysical coupling can reduce parameter errors, but only if the conceptual hydraulic model is correct. Considering an undisturbed soil core, *Mboh et al.* [2012a] also improved their parameter estimates when combining inflow measurements with ERT data-. They also highlighted the challenge of assigning appropriate weights to each data set in the objective function.

*Huisman et al.* [2010] combined soil moisture measurements with ERT monitoring data acquired over a dike built of uniform sand. They constructed a 2-D hydrological model and inverted for the homogeneous soil parameters of the dike using MCMC. They included the ERT data through a petrophysical relationship and inverted for unknown petrophysical parameters. They found that the permeability of the dike was the best-resolved parameter and that the combination of ERT and water content measurements reduced uncertainty.

*Mboh et al.* [2012b] used self-potential monitoring data acquired during infiltration and drainage experiments in a sand-filled column to invert for key parameters that describe the soil water retention and relative permeability functions. The hydrogeophysical coupling was obtained by linking the predicted streaming potential to the simulated water flow and water content distribution. While promising in the laboratory, the weak signal makes applications to field studies challenging [*Linde et al.*, 2011]. The petrophysical relationship needed to calculate the streaming potential at partial saturation is currently being debated [e.g., *Jougnot et al.*, 2012] and associated uncertainties are generally larger than for GPR and ERT.

### 7.4.2. Petrophysical coupling applied to transient groundwater processes

Another popular use of coupled hydrogeophysical inversion is for salt tracer tests in saturated aquifers. Dissolved salt increases the electrical conductivity of water, thereby decreasing electrical bulk resistivity. The changes in bulk resistivity can be sensed by ERT or EM methods and related to the fluid conductivity, which in turn can be linked to salinity. For moderate salinity, there is a linear relationship between salt concentration and fluid conductivity [e.g., *Keller and Frischknecht*, 1966], which can be calibrated for specific conditions.

Considering synthetic test cases, *Irving and Singha* [2010] and *Jardani et al.* [2013] both linked a flow and transport simulator at saturated conditions with an ERT forward solver to infer the permeability distribution. Both approaches used MCMC, but the parameterization was different. *Irving and Singha* [2010] used a fine spatial discretization, fixed the permeabilities of two facies and inverted for the probability of each cell to belong to one of the two facies. ERT data were inverted jointly with concentration data measured in boreholes and were found to mainly improve the estimates of the spatial correlation length. *Jardani et al.* [2013] used a pilot point parameterization to decrease the number of model parameters and inverted for the permeability at each pilot point. They included ERT, self-potential and tracer concentrations in the inversion and found that all three data sets contained valuable information on permeability.

*Pollock and Cirpka* [2010] considered a synthetic laboratory salt tracer experiment with ERT and hydraulic head measurements. They inverted for the permeability distribution using the mean arrival times of electrical potential perturbations and hydraulic head measurements. Their approach assumed a linear relationship between salt concentration and bulk electrical conductivity, but avoided the actual conversion between concentration and electrical conductivity. In a real sandbox experiment, *Pollock and Cirpka* [2012] recovered the detailed permeability structure. The full transient behavior was predicted very well even though only mean arrival times were used in the inversion. The inversion methodology developed by *Pollock and Cirpka* [2010, 2012] is very efficient as the forward problem that relates electrical potential difference perturbations to the saline tracer distribution and, hence, to the permeability field is formulated in terms of temporal moment-generating equations [*Harvey and Gorelick*, 1995]. Temporal moment-generating equations are widely used in hydrogeology and allow, for example, calculating the mean arrival time of a tracer by solving a steady-state equation instead of performing a full transient simulation.

*Johnson et al.* [2009] presented a data correlation approach where no specific petrophysical relationship was assumed, only the type of relationship (e.g., linear) between changes in fluid conductivity and changes in bulk electrical conductivity had to be chosen. They parameterized a 3-D subsurface model using pilot points and inverted for the permeability distribution using hydraulic head, fluid conductivity in 6 wells and surface ERT data. In their synthetic example (Fig. 7a), inversion of head and fluid conductivity data were unable to constrain the high-permeability zone (Fig. 7b). ERT data alone and especially joint inversion of hydraulic and ERT data greatly improved permeability estimates (see Fig. 7.5c and d). Due to the loose assumption of a correlation rather than a fixed petrophysical relationship between the hydrological state variable and the primary geophysical property, the approach of *Johnson et al.* [2009] could potentially be applied to a wide variety of applications.

*Kowalsky et al.* [2011] performed a coupled inversion of ERT and hydrogeochemical data to better understand the factors that influence flow and contaminant transport in a complex geological setting. They adapted their parameterization of the hydrogeological model at the field site to fit all available geophysical and geochemical data and to estimate permeability of the different model units, along with petrophysical parameters needed to use the ERT data. *Kowalsky et al.* [2011] were the first to link unsaturated and saturated flow and transport in a coupled hydrogeophysical inversion and their application highlights the complexity of real world problems. A parallel computing implementation of this approach [*Commer* *et al.*, 2014] will enable applications to even more complex environments and larger data sets.

*Dorn et al.* [2013] conditioned discrete fracture network realizations such that they were in agreement with field-based single-hole GPR reflection, tracer, hydraulic and televiewer data. The tracer test data were used to identify active fractures that intersected the borehole (inferred from electrical conductivity logs) and those within the formation (inferred from time-lapse GPR images that were sensitive to the tracer distribution). No petrophysical link was needed here, as it was the fracture geometry of the active fractures that was constrained by the GPR data. *Dorn et al.* [2013] stochastically generated three-dimensional discrete networks of active fractures and used a hierarchical rejection sampling method to create large sets of conditional realizations. Constraints offered by the GPR data made the stochastic scheme computationally feasible as they strongly reduced the set of possible prior models.

*Christiansen et al.* [2011] and *Herckenrath et al.* [2012] used field-based time-lapse gravity data to estimate porosity and permeability in shallow unconfined aquifers. They inverted hydraulic head measurements along with changes in the gravity response of the mass change associated with the fluctuating water table. *Christiansen et al.* [2011] inverted for the homogeneous porosity, permeability, evapotranspiration and riverbed conductance. They found that including gravity measurements significantly reduced parameter correlation between porosity and permeability and that especially porosity was better constrained when including gravity data. For a synthetic test case, *Herckenrath et al.* [2012] additionally included SNMR data in their coupled inversion for porosity and permeability.

*7.4.3 Petrophysical coupling applied to steady-state groundwater systems*

Calibration of groundwater models to data acquired under steady-state conditions is more challenging, due to the lack of dynamic forcing terms. Nevertheless, regional groundwater models are often built using hydraulic head distributions along with borehole descriptions and knowledge of the local geology. *Herckenrath et al.* [2013b] developed a joint hydrological inversion approach that included synthetic and field-based ERT and TDEM data to obtain a calibrated groundwater model. They used a combination of common interfaces and petrophysical relationships to link the geophysical response to the groundwater model. Using the same methodology, *Herckenrath et al.* [2013a] calibrated a saltwater intrusion model to field-based TDEM and hydraulic head data.

*Jardani et al.* [2009] linked hydraulic, thermal and self-potential simulators to jointly invert downhole temperature and self-potential data measured at the surface. In a geothermal field application, they built a model with 10 geological units. Using a stochastic inversion scheme, they inferred horizontal and vertical permeability of each model unit using borehole temperature data and self-potentials measured along a surface profile. The derived permeability values agreed well with those from hydrological studies.

*Straface et al.* [2011] jointly inverted field-based hydraulic head and self-potential measurements for the 3-D distribution of permeability in a shallow aquifer. To do so, they rely on the sensitivity of self-potential data to the water table level. Each model cell was first classified using a geostatistical multi-continuum approach and hydraulic inversion was performed on data collected during dipole pumping tests.

### 7.5. Outstanding challenges

Joint inversions of near-surface geophysical data and coupled hydrogeophysical inversions have advanced subsurface characterization, but important outstanding research challenges remain to be solved before their full potential can be reached.

### 7.5.1. Petrophysical relationships

When using petrophysical relationships within the inversion, it is necessary to first decide upon its functional form (e.g., is surface conductivity to be ignored when predicting electrical conductivity; is the petrophysical relationship based on volume-averaging or differential effective medium theory). The next step is to decide if petrophysical parameters are to be estimated within the inversion, assigned based on literature data, or estimated from independent laboratory or logging data. It is also essential to consider spatial variations of petrophysical parameters. For example, is it reasonable to assume that the cementation exponent in Archie-type relations is the same throughout a region consisting of different lithological units? This might be tempting as it simplifies the solution of the inverse problem, but it is often unlikely to be representative of reality.

Furthermore, is the petrophysical relationship assumed to be perfect (a one-to-one relation) or is it assumed to have uncertainty (almost always modeled with a Gaussian error model)? If it accounts for uncertainty, what is the spatial correlation structure of this error term? Should we assume that the error associated with the petrophysical relationship is the same throughout the model domain or that it is different at every grid cell? These questions are rarely addressed in near-surface geophysical or hydrogeophysical studies. Primary references concerning the most commonly used petrophysical relationships provide no or only very limited information about the spatial arrangement of the sample locations (the results are often presented in scatter plots).

To complicate things even further, most petrophysical relationships are originally defined at the scale of a representative elementary volume (REV), while model coupling is imposed at a much larger scale that is determined by the model parameterization and resolution of the geophysical or hydrological models. There are few reasons to believe that the same parameter values apply at both scales. This important topic has only been partially addressed in the hydrogeophysics and near-surface geophysics communities [*Day-Lewis and Lane*, 2004; *Moysey et al.*, 2005; *Singha and Moysey*, 2006].

The challenges discussed above are not specific to geophysics and arise in any application in which one state variable or property is linked to another one. For example, the relation between the water content and relative permeability or the relation between water content and capillary pressure, and so on. Nevertheless, ignoring the uncertainty related to the petrophysical relationship and its spatial dependence will often lead to overly optimistic uncertainty estimates that might overstate the information content in the geophysical data. How to properly account for these effects is an important challenge for future research.

### 7.5.2. Structural constraints

Structural approaches to joint inversion are popular as they avoid introducing explicit petrophysical relationships. The resulting inversion results are similar to those obtained from joint inversion through a known petrophysical relationship, but the structural approach is more robust when the petrophysical relationship is uncertain or spatially variable [*Moorkamp et al.*, 2011]. The cross-gradients constraints are used to obtain smooth models at a resolution that is much larger than the model discretization, and the gradients are calculated with respect to neighboring cells and are thus strongly mesh dependent. How do we assess if the underlying assumption of the cross-gradient function being zero is valid, or rather, for what applications and at what scale is the assumption reasonable? Can we formulate the cross-gradient constraints to operate at that scale? As we push the resolution limits of deterministic inversion, for example, by introducing full waveform inversion, it appears important to carefully assess if this type of constraint is meaningful at those finer scales. Furthermore, comparison with individual inversions, site-specific knowledge and petrophysical relationships are crucial to assess the validity of cross-gradients constraints for a given application.

### 7.5.3. Probabilistic vs. deterministic inversion

Probabilistic inversions, for example, based on MCMC are well suited to integrate multiple geophysical data sets and arbitrary petrophysical relationships [*Bosch*, 1999]. Uncertainty estimation is straightforward (but may be strongly affected by minor assumptions; e.g., *Linde and Vrugt* [2013]) and it is possible to perform the inversion within a modern geostatistical simulation context [*Hansen et al.*, 2012]. But, MCMC inversion is still out of reach for most applications with large data sets (millions of data points), high model dimensions (thousands of parameters), and advanced forward solvers (e.g., 3-D solvers). To apply MCMC, it is often necessary to reduce the number of unknowns, the number of data, and the accuracy of the forward solver. This implies that theoretically solid approaches to uncertainty assessments may become strongly biased and questionable [e.g., *Linde and Revil*, 2007].

The smoothness of Occam-style cross-gradients deterministic inversions helps to avoid excessive over-interpretation as it is clear from the outset that only an upscaled minimum-structure representation of the subsurface is sought. Such deterministic inversion results are often rather robust as the regularization term can be tweaked (by the inversion code or by hand) to get models that appear reasonable to the modeler. No such tweaking term should exist in probabilistic inversions, even if there is often a tendency to achieve a similar effect by manipulating the prior probability density function [*Scales and Sneider*, 1997].

Both the deterministic and probabilistic approaches have their merits. It is seldom possible to accurately state the actual prior information on model parameters and petrophysical relationships. It is thus often useful to consider multiple inversions with different underlying assumptions. It is a mistake to ignore uncertainty, but it might also be as dangerous to blindly believe in the uncertainty estimates provided by linear deterministic inverse theory or advanced probabilistic inversions. Current error models that incorporate the effects of the model parameterization, simplifications of the physics, the prior property fields, and petrophysical relationships are still far too simplistic. Improving these aspects and applying them to the joint inverse problem is an important future challenge for the near-surface geophysics and hydrogeophysics research communities.

### 7.6. Concluding remarks

Joint inversion can enhance model resolution and decrease interpretational ambiguities. Its application to near-surface geophysical data has become a robust and almost standard approach to integrate multiple geophysical data sets into consistent subsurface models at the highest possible resolution. The most robust coupling strategy is often structural (common layers, common lithological units or aligned gradients of property fields). Nevertheless, it is essential to carefully motivate a particular structural coupling approach for a given application to avoid the incorrect use of structural constraints and to identify when model coupling by petrophysical relationships is more suitable.

Joint inversion of geophysical and hydrological data has become very popular in recent years and the vast majority of publications rely on a coupled hydrogeophysical inversion framework. Hydrological models provide by simulation a predicted hydrological state that is transformed into a geophysical model through a petrophysical relationship. The misfit between the predicted geophysical forward response and the observed geophysical data is then used to guide the update of the hydrological model. This approach is often favored among hydrologists as the focus is on calibrating the hydrological model.

In field-based studies, it is useful to first identify incoherencies in data, geometry, and modeling results by performing individual inversions and comparing the results that are obtained with different methods. Blind application of joint inversion algorithms without careful assessments of data quality and coverage, model parameterization, the error model, geological setting and imposed prior or regularization constraints rarely leads to useful results. Future research avenues of interest include fully probabilistic joint inversions combined with complex prior models and improved statistical descriptions of petrophysical relationships and their scaling properties. These studies could also be most useful to better understand the fidelity of joint inversion results obtained within a deterministic framework for cases when a fully probabilistic treatment is computationally infeasible (e.g., most 3-D applications) and to understand which steps in the data acquisition, modeling and inversion process that are the most critical to ensure reliable results.

**Acknowledgements**

We are grateful for the constructive comments and suggestions provided by associate Editor Peter Lelièvre, Lee Slater and three anonymous reviewers.

**Figure captions:**

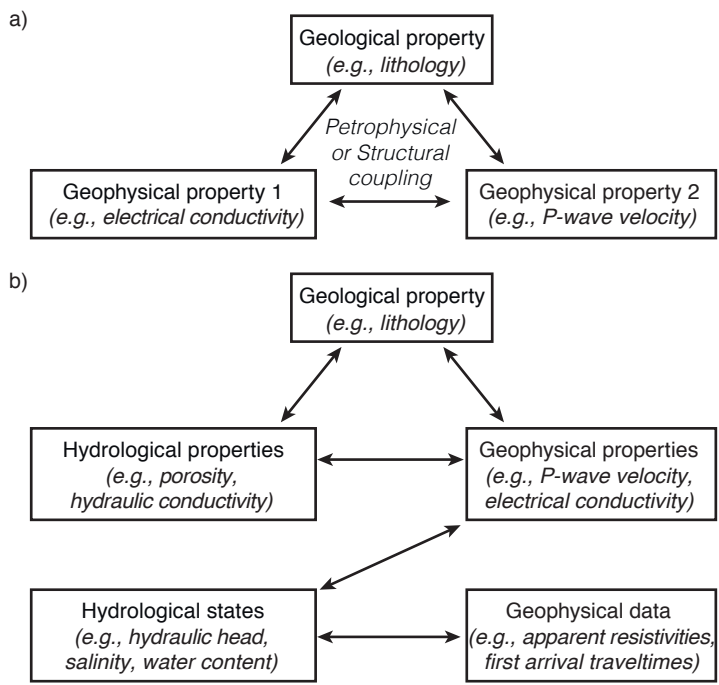

**Figure 7.1.** (a) Common coupling strategies for the joint inversion of near-surface geophysical data. (b) Joint inversion of hydrological and geophysical data opens up additional possibilities for model coupling. These schematic figures should not be interpreted as flow charts of joint inversion methods as they simply highlight the possible information exchange between multiple properties and data.

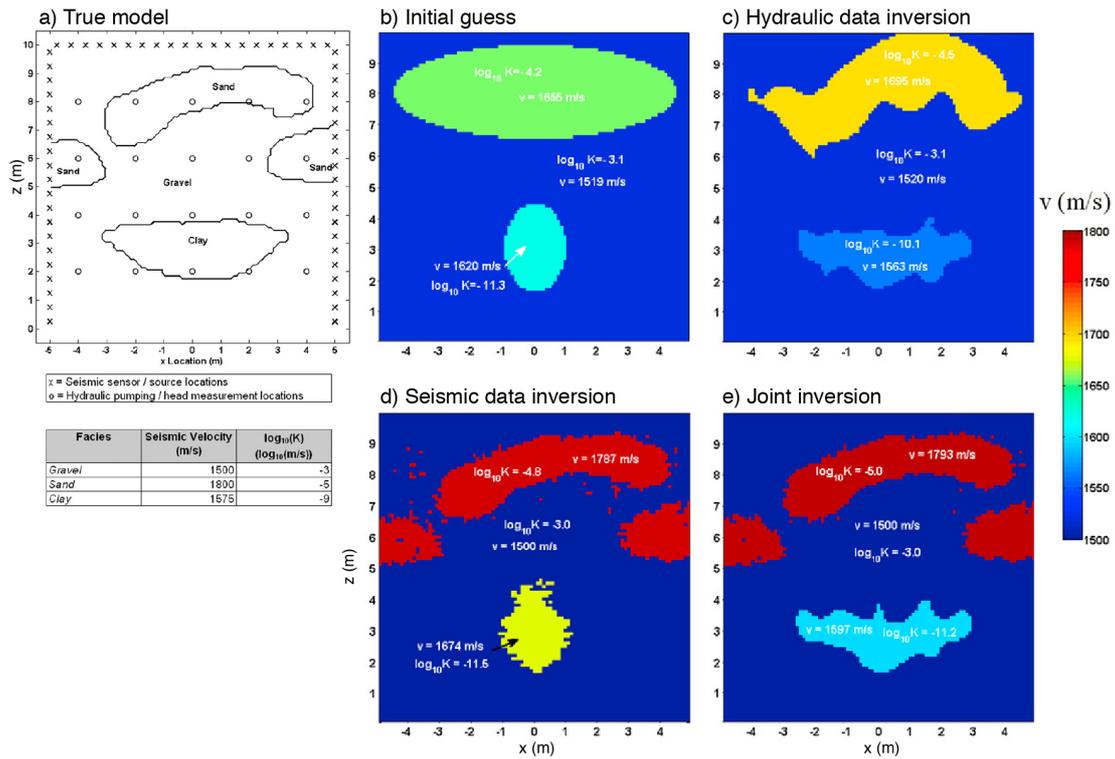

**Figure 7.2.** Results from zonal inversion using level sets, with the true model in (a), and the initial guess in (b). The individual inversion results of hydraulic and seismic data (c and d) are clearly improved in the joint inversion results (e). Modified from *Cardiff and Kitanidis* [2009].

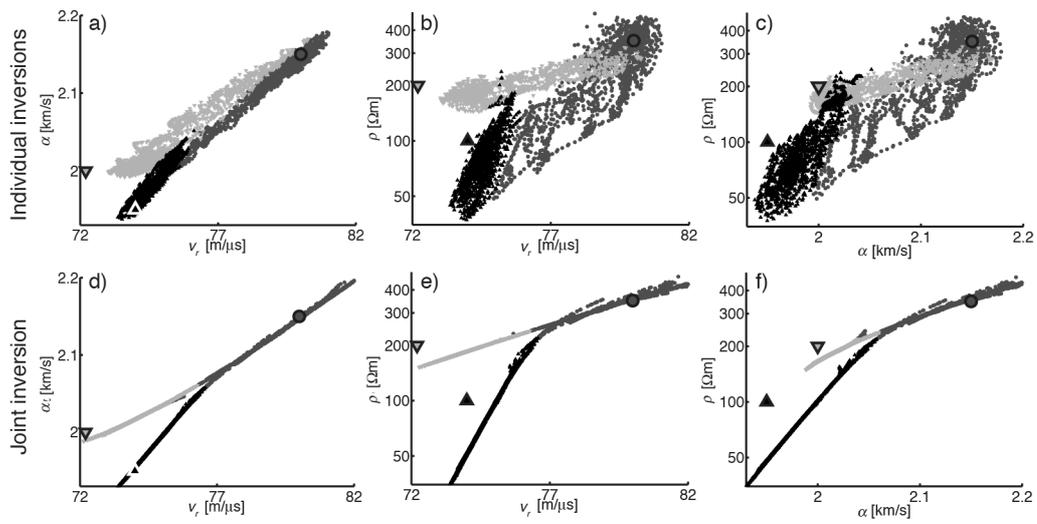

**Figure 7.3.** Cross-property plots of seismic velocity $\alpha$, GPR velocity $v_r$ and resistivity $\rho$. The diffuse scatter clouds from individual inversions (top) focus to clear linear features in the joint inversions (bottom). The joint inversion results are also closer to the zonal properties of the underlying synthetic lithological model (larger symbols). Modified from *Doetsch et al.* [2010b].

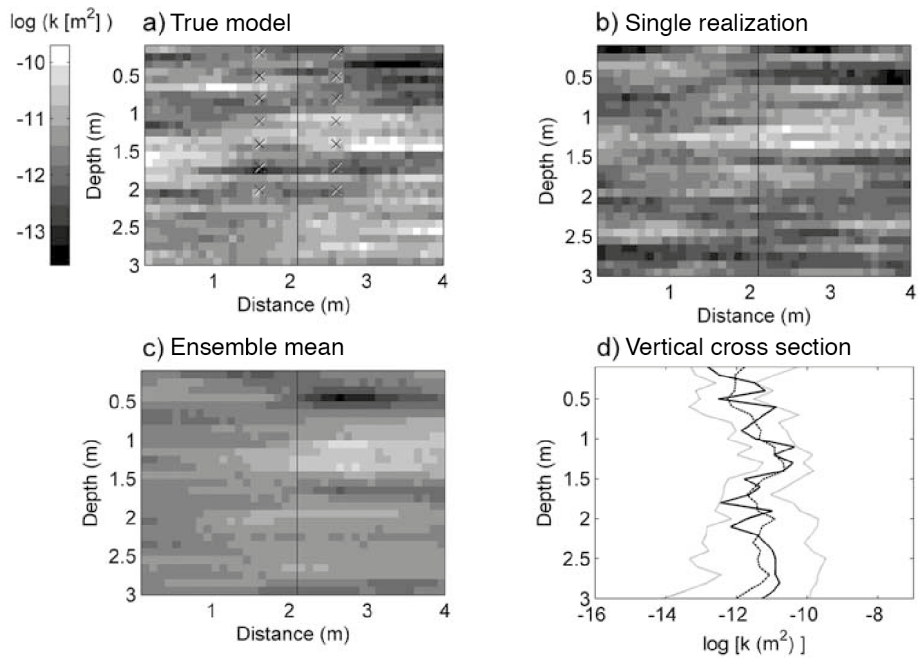

**Figure 7.4.** Input permeability distribution (a) and inversion results (b and c) from a synthetic tracer experiment. Both the single realization (b) and the ensemble mean (c) capture the main features, but the single realization shows more realistic variability. The vertical cross section (d) shows the true (solid line), mean (dashed line) and uncertainty bounds (gray lines) of permeability. From *Kowalsky et al.* [2005].

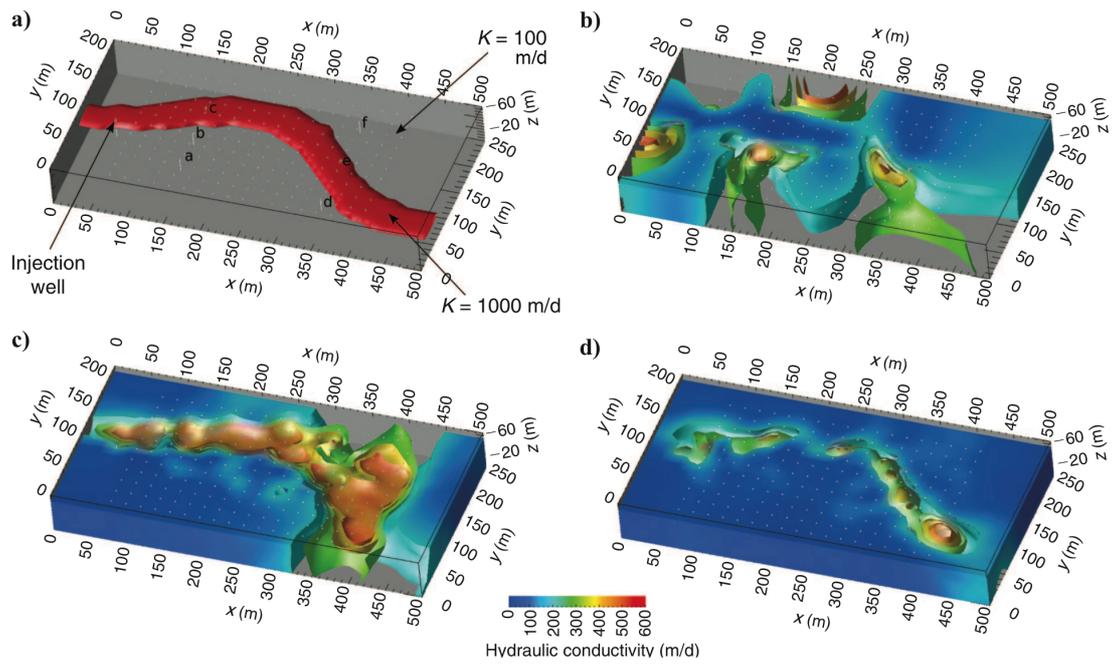

**Figure 7.5.** (a) True hydraulic conductivity distribution, (b) inversion of head and fluid conductivity data only, (c) ERT inversion results and (d) inversion of head, fluid conductivity and ERT data. From *Johnson et al.* [2009].